\documentclass[aps,prl,reprint]{revtex4-1}
\usepackage{epsfig}
\usepackage{graphicx}
\usepackage{amsmath}
\usepackage{amsfonts}
\usepackage{amssymb}
\usepackage{graphicx}
 \usepackage{wrapfig}
\usepackage[cp1251]{inputenc}
\usepackage{epsfig}
\usepackage{graphicx}
\usepackage{amsmath}
\usepackage{amsfonts}
\usepackage{amssymb}
\usepackage{graphicx}
 \usepackage{wrapfig}
\usepackage{textcomp}
\usepackage{pgf,pgfarrows,pgfnodes,pgfautomata,pgfheaps,pgfshade}
\usepackage{amsmath,amssymb}
\usepackage[config]{caption,subfig}
\usepackage[colorlinks=true, linkcolor=blue]{hyperref}
\usepackage{marginnote}  % margin notes package
\newcommand{{\sign}}{\rm sign}
\newcommand{{\const}}{\rm const}

\begin{document}
\title{Role of qubit-cavity entanglement for switching dynamics of quantum interfaces in superconductor metamaterials}
\author{Sergey V. Remizov$^{1,2}$}
\author{Dmitriy S. Shapiro$^{1,2,3}$}
\email{shapiro.dima@gmail.com}
\author{Alexey N. Rubtsov$^{1,3,4}$}
\affiliation{$^1$Dukhov  Research Institute of Automatics (VNIIA),
127055 Moscow, Russia}
\affiliation{$^2$V. A. Kotel'nikov Institute of Radio Engineering and Electronics, Russian Academy of Sciences, 125009 Moscow, Russia}
\affiliation{$^3$National University of Science and Technology MISIS, Leninsky prosp. 4, Moscow, 119049, Russia}
\affiliation{$^4$Russian Quantum Center,   Skolkovo, 143025 Moscow Region, Russia}

\begin{abstract}
	
We study quantum effects of strong driving field  applied to dissipative  hybrid qubit-cavity system which are relevant for a realization of quantum gates in superconducting quantum  metamaterials. We demonstrate that  effects of strong and  non-stationary drivings  have significantly quantum nature and can not be treated by means of mean-field approximation.
This is shown from a comparison of steady state solution of the standard Maxwell-Bloch equations and   numerical solution of Lindblad equation on a  density matrix. We   show that mean-field approach provides very good agreement with the density matrix    solution   at not very strong drivings $f<f^*$ but at $f>f^*$  a growing value of quantum  correlations between fluctuations in qubit and photon sectors changes a behavior of the system.  We show that in regime of non-adiabatic  switching on of the driving such a quantum correlations influence a dynamics of qubit and photons  even at weak $f$.

\end{abstract}

\maketitle

\subsection{Introduction}
Quantum metamaterials are hybrid systems consisting of arrays of qubits coupled to the photon modes of a cavity \cite{Macha,Astafiev,Rakhmanov, Fistul,ZKF,SMRU,Brandes,Zou}. In solid state structures the qubits are realized using  nitrogen-vacancy (NV) centers in diamonds \cite{nv-centers-0,nv-centers, nv-centers-1}, and  spins of $^{31}$P dopants  in $^{28}$Si crystals \cite{Morton} or Cr$^{3+}$  in Al$_2$O$_3$ samples \cite{Schuster}, and   superconducting Josephson qubits  \cite{MSS, Orlando,mooij}. Among of others, the  Josephson qubits are particularly perspective for an implementation of quantum gates  \cite{DiCarlo, MSS, Nation,  Clarke} due to their high degree of   tunability. Frequency of excitation, given by an energy  difference between ground and excited states, can be controllably tuned in a wide range using the external magnetic flux threading a loop of the qubit. Modern technology allows for a production of metematerial structures obeying sophisticated geometry and low decoherence effects.

High nonlinearity of the qubit excitation spectrum, combined with low decoherence, gives rise to unusual properties of quantum metamaterials, distinguishing them from the linear-optical metastructures. These unusual features are associated with intrinsic quantum dynamics of qubits and photon degrees of freedom. They are revealed in the optical response of a metamaterial to the external strong pump field, driving the system away from its ground state. A textbook example is the rotation on a Bloch sphere of the state of a single qubit subjected to an external field pulse. The well-understood solution for dynamics of a single qubit is commonly used as a key building block in the mean-field description of complex metamaterials containing a number of qubits and cavity modes.
Assuming no correlations between the qubits and photons, one comes to the set of Maxwell-Bloch equations virtually describing qubits coupled to a classical field of the cavity and (or) external pump.

This article is devoted to the role of quantum entanglement between qubit and cavity modes of the superconducting metamaterial. Whereas it is generally clear that these correlation effects beyond Maxwell-Bloch scheme are revealed in strong-driving regimes, their quantitative role in an experimentally/technologically relevant situation is not yet studied. At the same time, such study is highly motivated by the quantum technology development, because
a realization of qubit gates and operation of quantum simulators  assume applying of driving fields of strengths comparable with qubit-cavity coupling energy.  We argue that a quantitative description of an operation of realistic quantum metamaterials, which involve non-adiabatic and strong perturbations, assumes taking into account quantum correlations.

We present a study of the a simple yet realistic model of the quantum interface, defined as
a dissipative hybrid system containing a resonant qubit being connected to the cavity mode and simultaneously subjected to the strong external  field. We assume that the  two-level system is highly anharmonic  flux qubit, being a loop with several Josephson junctions, where highest levels are not excited by the external drivings.
 Hybridization between the qubit and the cavity mode provides a transfer
of the pump photons to the cavity mode via qubit excitations. Therefore the internal qubit dynamics is fingerprinted in the cavity field, and can be later read out or transferred to another qubit.
We describe the evolution of the many-body density matrix of the system using the Lindblad equation, and compare the results with whose obtained using Maxwell-Bloch approximation. We observe, that for a constant driving field the two approaches give almost same results (that is, qubit-cavity correlations are negligible) up to
certain threshold value of the pump $f^*$   depending mainly on relaxation rates in a cavity and in a qubit.  For higher pump field, the effect of correlations rapidly grow, making Maxwell-Bloch approximation quite inaccurate.  There is a remarkable artifact which follows from the Maxwell-Bloch approach, but is not present in the many-body description:  a hysteresis in photon number as a function of $f$. This behavior shows up in a certain range around of the threshold $f^*$ if a coupling energy between photons and the qubit is large enough. Furthermore, we find that non-adiabatic switching  on  of the driving, from zero to a certain value, reveals discrepancy between mean-field and exact solutions even for drivings below the steady-state threshold $f^*$.

%In this case an essential role is played by strongly non-linear effects generated by to two-level nature of qubit, when it is brought into a superposition state, and and also by quantum correlations between qubit and a photon field. In this work we analyze  these effects for a  dissipative hybrid system of resonant qubit and cavity mode.
%We find out that, first, in steady state regime the standard mean-field approach based on Maxwell-Bloch equations provides very good agreement with exact numerical solution of Lindblad equation at not very strong drivings. At strong drivings Maxwell-Bloch equations gives a discrepancy, especially for photon field. The difference between Maxwell-Bloch and exact solution at strong drivings results from a growing of correlations between fluctuations of qubit and photons degrees of freedom.

\section{Theoretical approach}
 \label{theory}

Description of circuit quantum electrodynamics of  superconducting metamaterials of qubits and transmission line are  reduced to a Hamiltonian of Tavis-Cummings model \cite{Carmichael}. In our analysis we start from more simple situation of a single qubit which is coupled to photon mode. Quantum mechanical description is reduced to well-known Janes-Cummings model which is exactly integrable. Namely, Hamiltonian of an isolated qubit-cavity system is
$$H_{JC}=\omega_R a^+a +     \epsilon  \sigma^+ \sigma^-  +g  (a\sigma^+ + a^+\sigma^-).$$
First term describes excitations in photon mode of the resonator, where bosonic operators $a,a^+$ obey commutation rules $[a,a^+]=1$. Second term is related to excitations in qubit where $\sigma^+ , \sigma^- $ are Pauli operators.
External transversal driving applied to a qubit is accounted for by
$$
H_{ext}=\frac{f(t)}{2} \left(e^{-i\omega t}\sigma^+ + e^{i\omega t}\sigma^-\right)
$$
where $f(t)$ is slow envelope function and $\omega$ is fast reference frequency. In our studies we are limited by single non-adiabatic switching of the form
\begin{equation}
f(t)=f\theta(t). \label{f-t}
\end{equation}
The system under consideration is coupled to a dissipative environment, hence, we employ Lindblad equation on the  density matrix $\rho(t)$ dynamics written in  many-body basis of qubit and photon states.   The Lindblad equation reads as
\begin{equation}
i(\partial_t\rho(t) - \Gamma[\rho(t)])=[H(t),\rho(t)]. \label{lindblad}
\end{equation}
where full Hamiltonian
$$
 H(t)=H_{JC}+H_{ext}(t)
$$
and relaxations in qubit and cavity are taken into account by means of
\begin{multline}
\Gamma[\rho]=\frac{\kappa}{2}(2a\rho a^+ - a^+a \rho- \rho a^+a)
+ \\
+\frac{\Gamma_1}{2} (2\sigma^-\rho\sigma^+ - \sigma^+\sigma^-\rho- \rho\sigma^+\sigma^-).
\end{multline}

In our   numerical solution we calculate $\rho(t)$ by means of direct integration of Lindblad equation   in a  truncated Hilbert space. In approximate mean-field techniques we derive equations on averages from this equation (\ref{lindblad}) as well.

Note that everywhere below we perform transition into rotating frame basis related to the main frequency $\omega$ of driving signal which is tuned in resonance with cavity mode frequency $\omega=\omega_R$.
The full Hamiltonian $H(t)$ in (\ref{lindblad}), given by $H_{JC}+H_{ext}(t)$,  in this $\omega$-rotating frame basis reads
\begin{equation}
H(t)= \Delta \sigma^+\sigma^- +g (a\sigma^+ + a^+\sigma^-)+\frac{f(t)}{2} \sigma^x. \label{h}
\end{equation}
where qubit detuning is $\Delta=\epsilon-\omega_R$.

  From the Lindblad equation
$ i(\partial_t\rho(t) -\Gamma[\rho(t)])=[H(t),\rho(t)]$
we derive  equations for averages $\langle a \rangle, \langle a^+ \rangle, \langle \sigma^{\pm} \rangle,\langle \sigma^z \rangle$. This is done with use of definitions, e.g. applied to $a$, by the following scheme
\begin{equation}
\partial_t \langle  a (t) \rangle ={\rm Tr}( \partial_t\rho(t) a)={\rm Tr}( -i[H(t),\rho(t)]a-\Gamma[\rho(t)])a ).
\end{equation}
We apply the following standard mean-field  approximation: we  factorize the averages     \begin{equation}
\langle a\sigma^+ \rangle\rightarrow \langle a\rangle \langle \sigma^+ \rangle, \langle a\sigma^z \rangle\rightarrow \langle a\rangle \langle \sigma^z \rangle \label{factorization-m-b}
\end{equation}
which appears in r.h.s. parts of equations on $\langle a \rangle, \langle a^+ \rangle, \langle \sigma^{\pm} \rangle,\langle \sigma^z \rangle$. On the level of density matrix this corresponds to the introduction of the reduced density matrices $\rho_q$ and $\rho_{ph}$ and the full one $\rho_{mf}=\rho_q\otimes \rho_{ph}$.
This factorization (\ref{factorization-m-b}) is an approximation where we neglect correlation between  fluctuations in qubit and photon mode ($\delta\sigma^{\pm}, \delta\sigma^{z}$ and $\delta a, \delta a^+$)
\begin{equation}
\langle a\sigma^{z,\pm} \rangle=\langle a\rangle \langle \sigma^{z,\pm} \rangle+\langle \delta a\delta\sigma^{z,\pm}\rangle. \label{fluctuations-0}
\end{equation}

After such the  factorization (\ref{factorization-m-b}) we end up with Maxwell-Bloch  non-linear equations  (we do not write $\langle \rangle$ for brevity)
%$$
%\dot n(t)=-\kappa  n(t)+i g \sum _{j=1}^{\text{Nqubits}} (a(t) %\text{\sigma^+}(t)-\text{a^+}(t) \text{(sigma_j^-(k)(t))}
%$$
%\begin{equation}
%\partial_t  n (t)  = -\kappa a(t) + i g\sum\limits_{j=1}^{N_q} \left(\sigma^+(t)a(t) - %\sigma^-(t)a^+(t)  \right),
%\end{equation}
\begin{equation}
\partial_t a (t)  = -\frac{\kappa}{2}a(t)-i g  \sigma^-(t), \quad c.c., \label{a}
\end{equation}
\begin{equation}
\partial_t    \sigma^+  (t)  = (i\Delta-\Gamma_1/2) \sigma^+  (t) -i\left(\frac{f(t)}{2}+g a^+(t)\right) \sigma^z(t),   c.c., \label{sigma_minus}
\end{equation}
\begin{multline}
\partial_t  \sigma^z (t)  =- \Gamma_1  \left(\sigma^z(t)+1\right)  + \\ +2 i g \left( a^+(t)\sigma^-(t)-a(t)\sigma^+(t)\right)+\\ +i f(t) \left(\sigma^-(t)-\sigma^+(t)\right). \label{sigma-z}
\end{multline}
Note, that photon number dynamics is found from solution for $a(t)$ in such a mean-field technique as
\begin{equation}
n_{ph} (t)  = |a(t)|^2. \label{n-mb-0}
\end{equation}

\section{Results}
\subsection{ Steady state regime}
In this part of the paper we demonstrate a comparison between results  obtained from solutions of Lindblad (full many-body density matrix)  and Maxwell-Bloch (mean-field) equations in a wide range of drivings $f$. Here and below we are limited by fully resonant regime where $\epsilon=\omega_R$, i.e. the detuning is zero $\Delta=0$.  We evaluate numerically the $f$-dependences for qubit excited state  occupation number $n_q$, generated photon number $n$ in the cavity and correlators $\langle \delta a\delta\sigma^{z,\pm}\rangle$. All the data presented in the paper are obtained for the system with the parameters $\Gamma_1=0.5$ MHz, $\kappa=0.4$ MHz. This section is devoted to the steady-state regime emerging after a long evolution of the system subjected to the driving field having a constant amplitude and phase.

\begin{figure}[h]	\includegraphics[width=\linewidth]{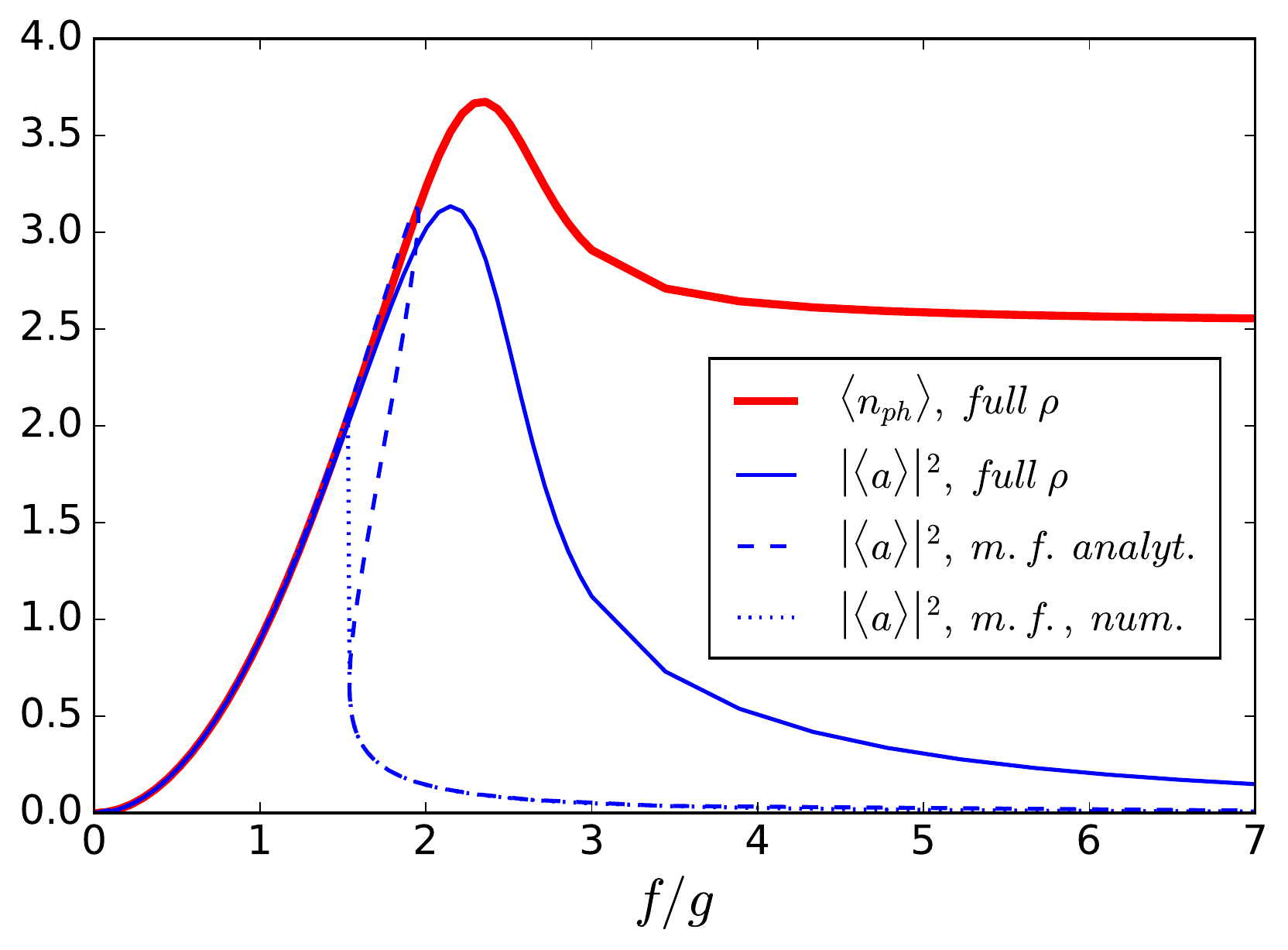}
	\caption{Photon number vs driving amplitude $f$ in the steady state regime.} \label{result:stationary:nph}
\end{figure}

\begin{figure}[h]
	\includegraphics[width=\linewidth]{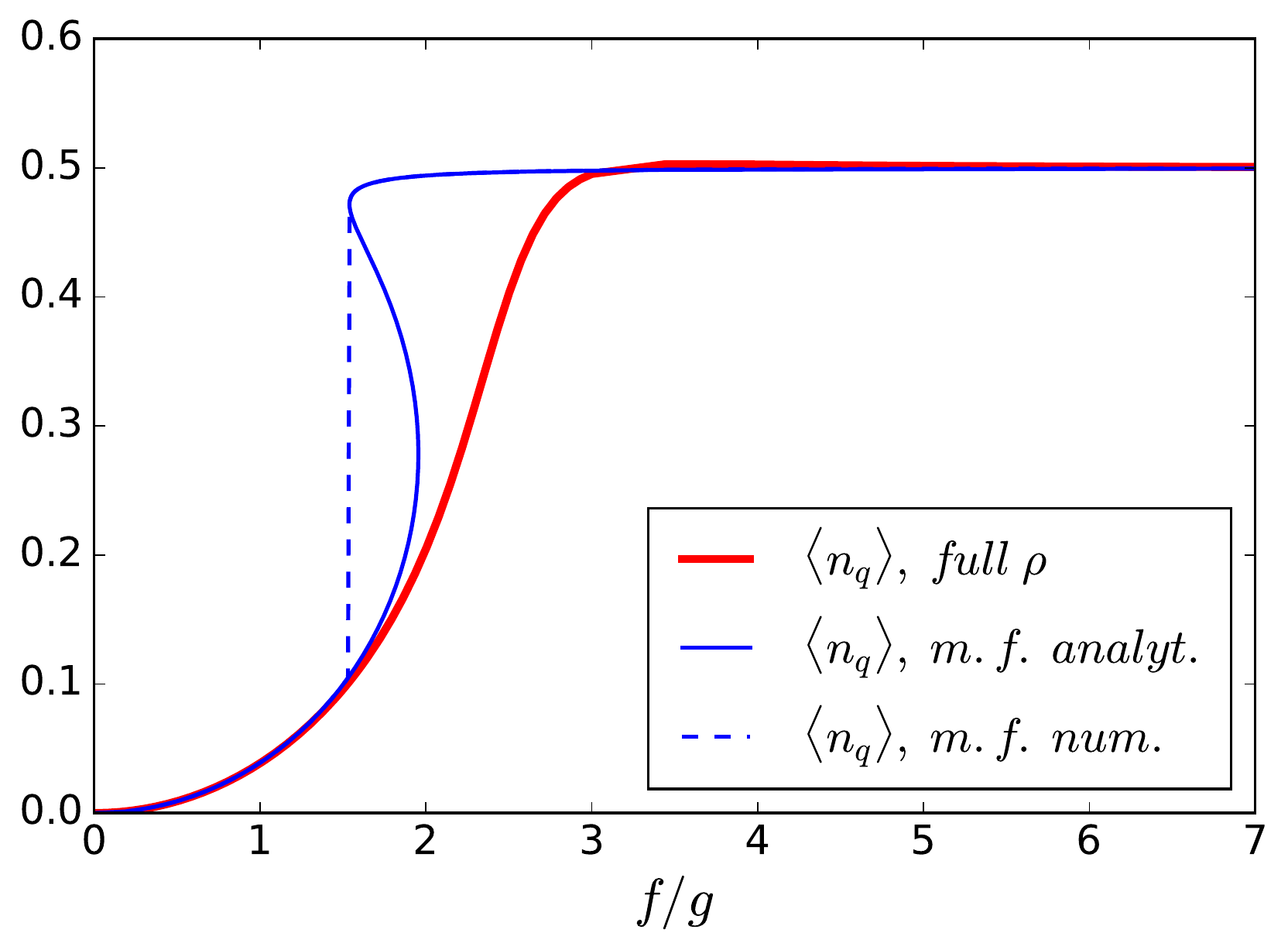}
	\caption{Qubit occupation number vs driving amplitude $f$ in the steady state regime} \label{result:stationary:nq}
\end{figure}

We observe from Figures \ref{result:stationary:nph} and \ref{result:stationary:nq} a very good agreement between the mean-field and the full density matrix  solutions for $n_{ph}$ and $n_q$ at $f<f^*$ where the value of $f^*$ divides a ranges of weak and strong field steady state regimes. At $f> f^*$ we observe an agreement for qubit occupation number which is $n_q=1/2$ in both of solutions. Indeed there are significant distinctions in behavior of photon degree of freedom: in strong field  limit of $f> f^*$  the photon number $n$ decays to zero in Maxwell-Bloch solution but saturates to a finite value in the Lindblad numerical calculation. 

The steady state solution of Maxwell-Bloch equations can be analyzed to explain the observed differences. 
 Taken the l.h.s. parts of the equations (\ref{a},\ref{sigma_minus}) and their conjugates equal  to zero, the following
  relations between $\langle a \rangle, \langle a^+ \rangle, \langle \sigma^{\pm} \rangle$ and $n_q=(\langle \sigma^z \rangle +1)/2$ are derived
\begin{equation}
\left(
\begin{array}{c}
\langle \sigma^{-} \rangle \\ \\
\langle \sigma^{+} \rangle \\ \\
\langle a \rangle \\ \\
\langle a^+ \rangle \\
\end{array}
\right)=
\left(
\begin{array}{c}
	-\frac{i f (2n_q-1)    \kappa }{4 g^2 (2n_q-1)   -\Gamma_1  \kappa } \\ \\
	\frac{i f (2n_q-1)    \kappa }{4 g^2 (2n_q-1)   -\Gamma_1  \kappa } \\ \\
	-\frac{2 f g (2n_q-1)   }{4 g^2 (2n_q-1)  -\Gamma  \kappa } \\ \\
	-\frac{2 f g (2n_q-1)   }{4 g^2 (2n_q-1)   -\Gamma_1  \kappa } \\
\end{array}
\right). \label{a-sigma-sol}
\end{equation}
Combining these results with (\ref{sigma-z}) with zero l.h.s. part we obtain the relation between $n_{ph}$ and $n_q$  
\begin{equation}
n_{ph}=-4n_q(2n_q-1)\frac{ g^2}{\kappa^2}.
\label{nph}
\end{equation}	
 The relation between qubit occupation number itself and driving amplitude $f$ is given by the  implicit expression which can be found from (\ref{a-sigma-sol}) as well
 \begin{equation}
 f=\frac{|\Gamma_1 \kappa - 4 g^2 (2n_q-1)|}{\kappa}\sqrt{\frac{ n_q}{1-2n_q}}. \label{f}
 \end{equation}	

Definitely,  the zero value of $n_{ph}$  resulting from Eq. (\ref{nph}) at large $f$, when qubit occupation number is saturated to  $n_q=1/2$ (see Fig. \ref{result:stationary:nq}), is wrong. A correct value for $n_{ph}$ can be easily found from the Hamiltonian (\ref{h}) in the limit of $f\gg g$. Namely, qubit ground state in such a limit is odd superposition $|\psi_{gs}\rangle=(|g\rangle-|e\rangle)/\sqrt{2}$, and, hence, $\sigma^\pm=1/2$. After that, we find  perturbatively steady state $a=-2i(g/\kappa) \sigma^-$  from (\ref{a}) for the dissipative system, yielding $n_{ph}=|a|^2=(g/\kappa)^2$ from the mean-field definition of $n_{ph}$ (\ref{n-mb-0}). This result is in agreement with the tendency to saturation of photon number $n_{ph}$ at large $f$ observed in the numerical solution.

Other comment is about the bistability region in the Maxwell-Bloch result seen in Figure \ref{result:stationary:nph} and \ref{result:stationary:nq}.  Mathematically it is due to the fact that  (\ref{f}) is a 3-rd order equation with respect to $f$. There exists a range for $f$, where three solutions for $n_q$, and, consequently, three values of $n_{ph}$ at a given $f$ are possible.  The condition for an existence of the three solutions in Maxwell-Bloch equations in this stationary regime is
	\begin{equation}
	g>\sqrt{2 \Gamma_1 \kappa}. \label{gc}
	\end{equation}
This condition follows from the expression for two extrema  of the inverse relation between $n_q$ and $f$ (shown as dased curve in Fig. \ref{result:stationary:nq}): $$n_{q}^{(1,2)}= \frac{1}{8}\left( 3\pm \sqrt{1-\frac{2  \Gamma \kappa}{  g^2}} \right).$$ One of the three solutions appears unstable and does not show up in the curves obtained numerically. The two others are stable and give rise to a bistability regime similar to the one in \cite{SavageCarmichael}  where a driving was applied to photon mode.  We insist, however, that the solution of the Lindblad equation for the many-body density matrix does not contain such a bistable regime and we therefore interpret it as an artifact of the mean-field approximation.

\begin{figure}[h]
	\includegraphics[width=\linewidth]{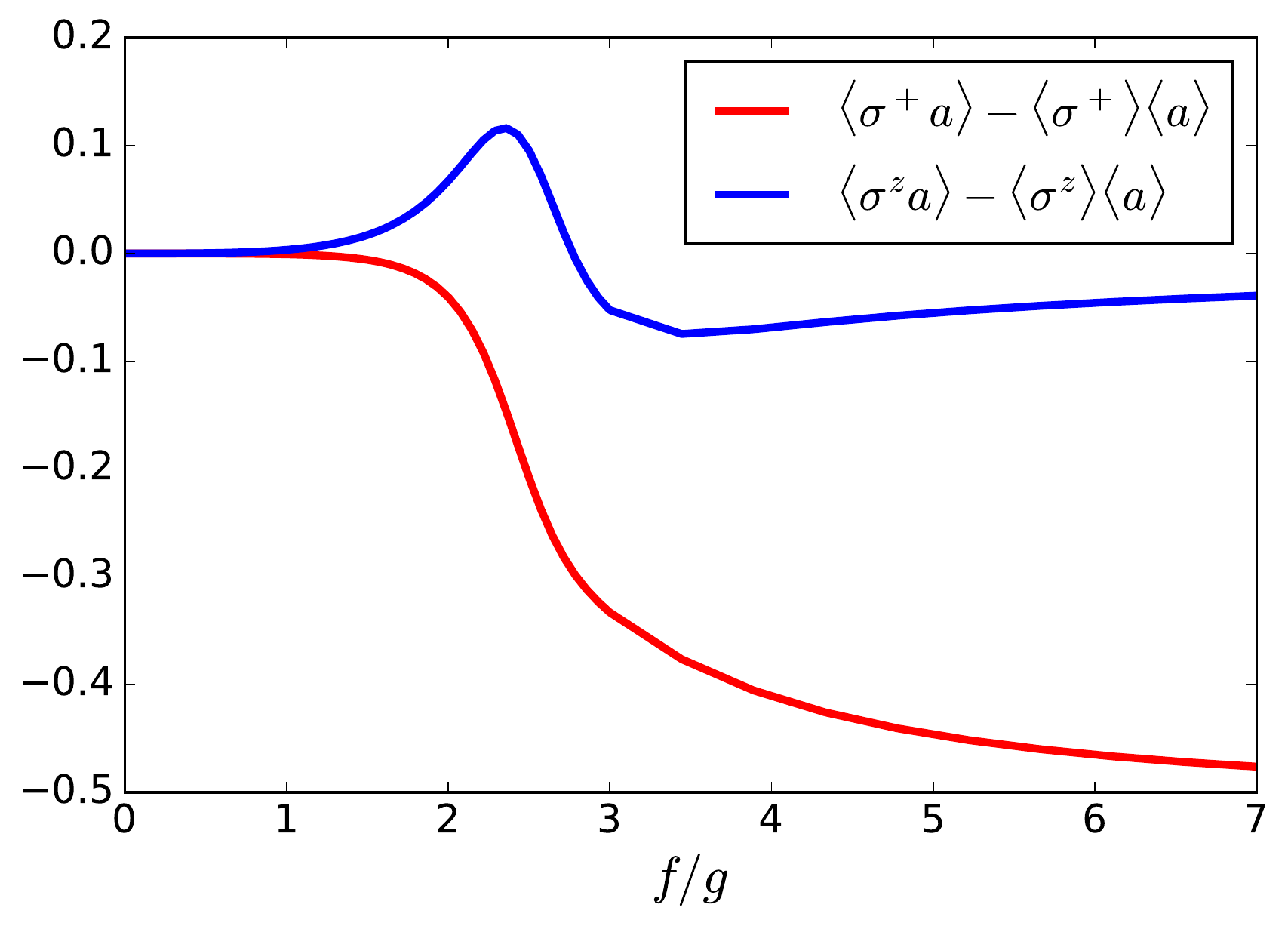}
	\caption{Correlator of fluctuations $ \langle \delta a\delta\sigma^{+}\rangle$ and $ \langle \delta a\delta\sigma^{z}\rangle$ extracted from solution of the Lindblad equation  for the full density matrix.} \label{result:stationary:spl-a}
\end{figure}

It is important that non-zero  correlators $\langle \delta a\delta\sigma^{z,\pm}\rangle$ demonstrate the increase of the effect of quantum  fluctuations in the regime of strong driving $f>f^*$, see Fig. \ref{result:stationary:spl-a}.
In the regime of strong coupling (\ref{gc}) the typical $f^*$ can be estimated from the mean-field relation (\ref{f}) as follows
$$
f^*\sim {\rm max}[\Gamma_1,\frac{g^2}{\kappa}].
$$
 As it is seen from the curve for $n_{ph}$ these fluctuations make a significant contribution in the photon sector of the system. Value of the fluctuations can be perturbatively estimated from the Maxwell-Bloch equations :
 $$
 \langle \delta a\delta\sigma^+\rangle=\frac{2i\kappa g (2n_q-1)(2g n_{ph}+f a)}{(2\kappa+\Gamma_1)\Gamma_1} - \frac{2 i g n_q}{2\kappa+\Gamma_1}
 $$
This correlator saturates to a non-zero value of $$\langle \delta a\delta\sigma^+\rangle_{f\gg f^*} = \frac{-i g}{2\kappa+\Gamma_1}$$ in the limit of strong driving where the qubit occupation number is $n_q=1/2$. In the Figure  \ref{result:stationary:spl-a} we present the results for the correlators obtained from the Lindblad solution for the full density matrix. The saturation of $\langle \delta a\delta\sigma^+\rangle$ at high $f$, appeared in the mean-field approach, is observed in these data as well.

\begin{figure}[h]
	\includegraphics[width=\linewidth]{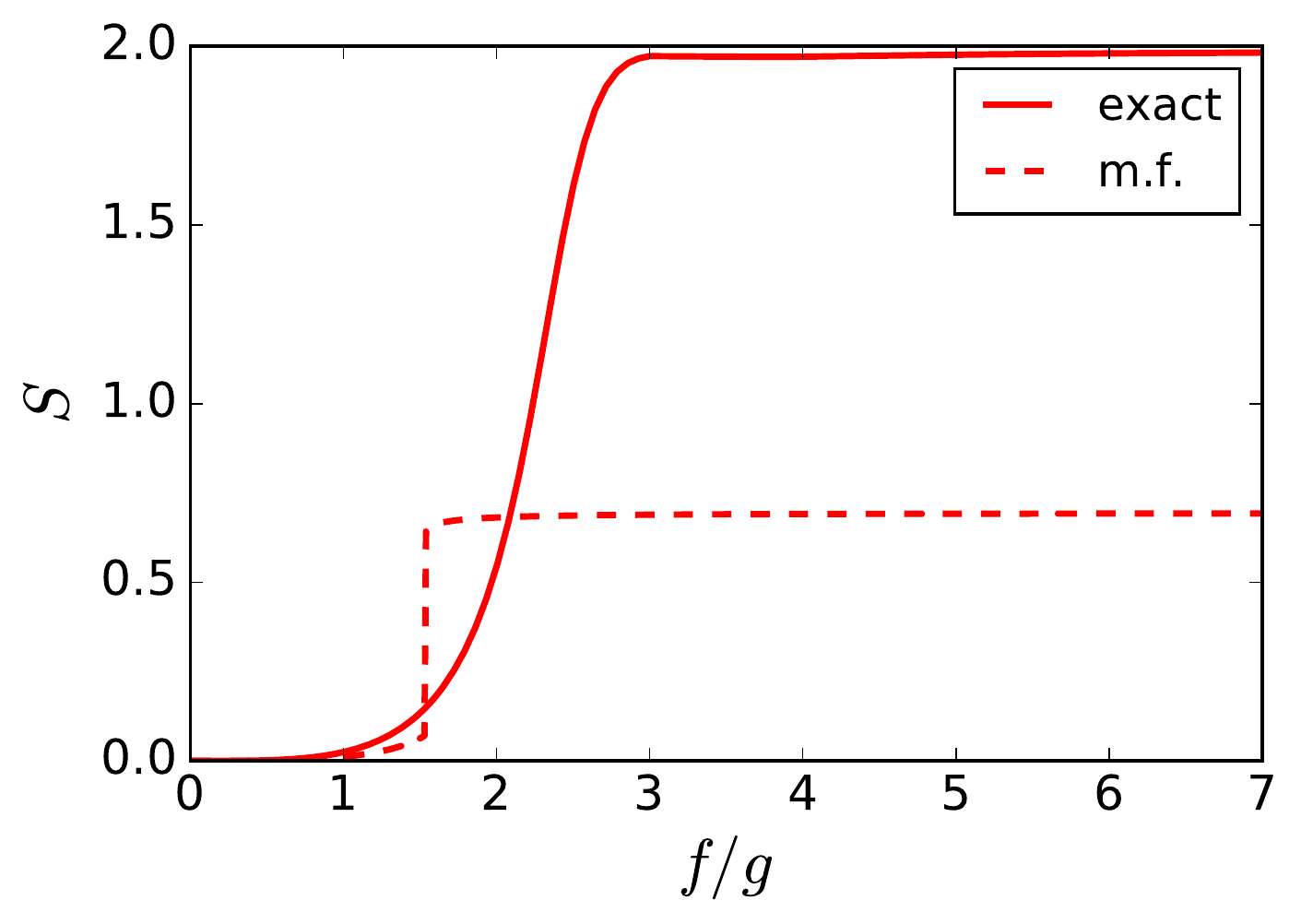}
	\caption{Entropy vs driving amplitude $f$ in the stationary regime. The solid curve is related to the density matrix  found from the solution of Lindblad equation. The dashed curve describes entropy calculated within the mean-field approximation. } \label{result:stationary:s}
\end{figure}

In Figure \ref{result:stationary:s} we show the numerical results for the von Neumann entropy $S=-{\rm Tr}\rho \ln \rho$. The solid curve demonstrates $S(f)$ calculated from the Lindblad approach while the dashed one is related to the mean-field approximation where the effective Hamiltonian include the values of $\langle a \rangle, \langle a^+ \rangle, \langle \sigma^{\pm} \rangle$ found from the solution of  Maxwell-Bloch equations. The difference between them at $f>f^*$ shows again that there is a significant entanglement between the qubit and photon degrees of freedom in the strong driving domain. The mean-field solution assumes that the many-body density matrix is a direct product of the qubit and photon ones $\rho_{mf}=\rho_{ph} \otimes \rho_q$, where the elements responsible for the entanglement are zero. These non-diagonal elements of the density matrix, taken into account in the solution of the  Lindblad equation,  increase the entropy.

\subsection{Non-stationary regime}

\begin{figure}[h]	\includegraphics[width=\linewidth]{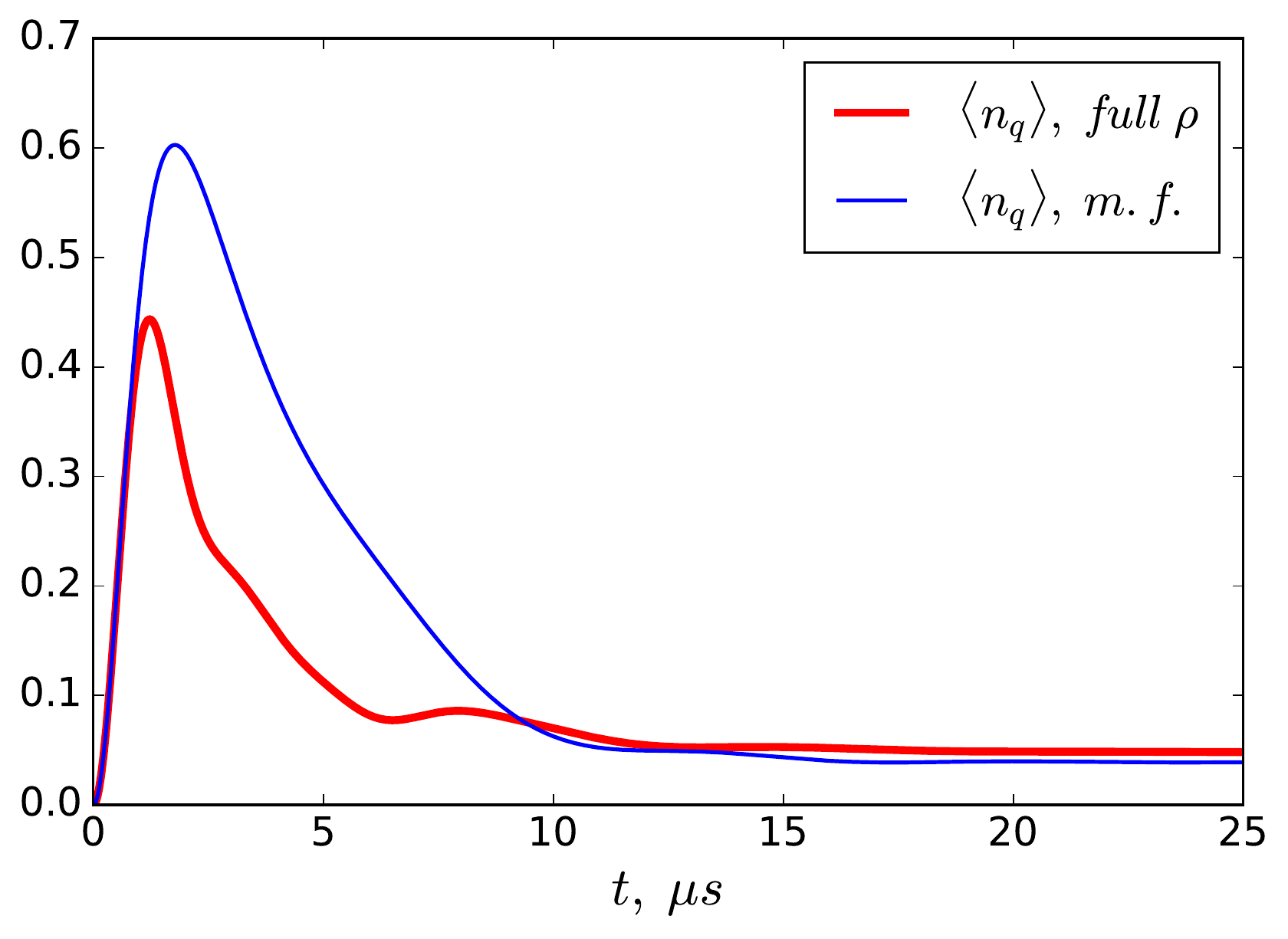}
	\caption{Time evolution of the qubit occupation number $n_q(t)$ found from the solution on the full density matrix and the mean-field approach at $f=g$.} \label{result:nonstationary:ns}
\end{figure}

\begin{figure}[h]	\includegraphics[width=\linewidth]{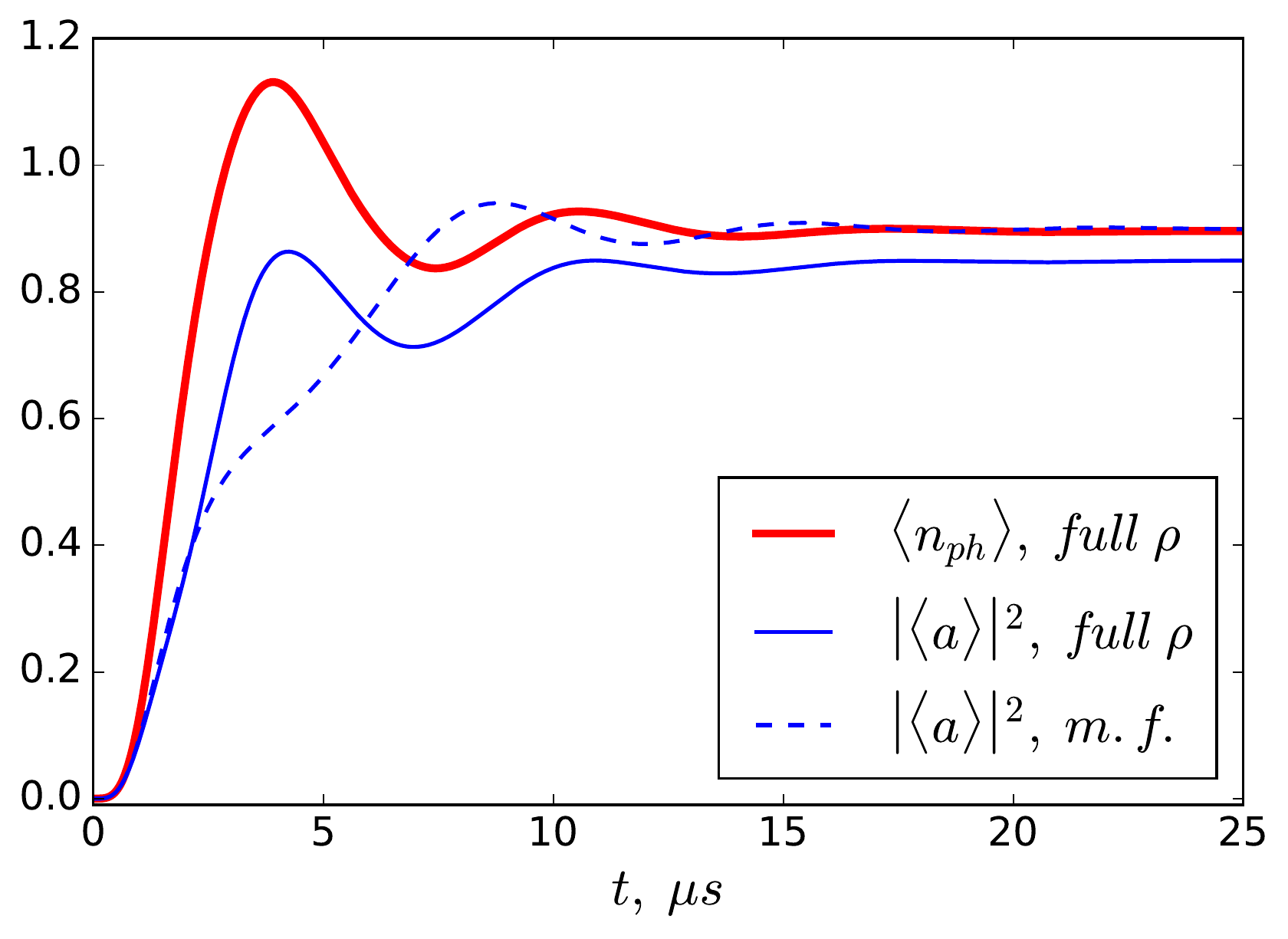}
	\caption{Time evolution of photon occupation number $n_{ph}(t)$ found from solution of the Lindblad equation  on the full density matrix and the mean-field approach at $f=g$. } \label{result:nonstationary:nph}
\end{figure}

\begin{figure}[h]
	\includegraphics[width=\linewidth]{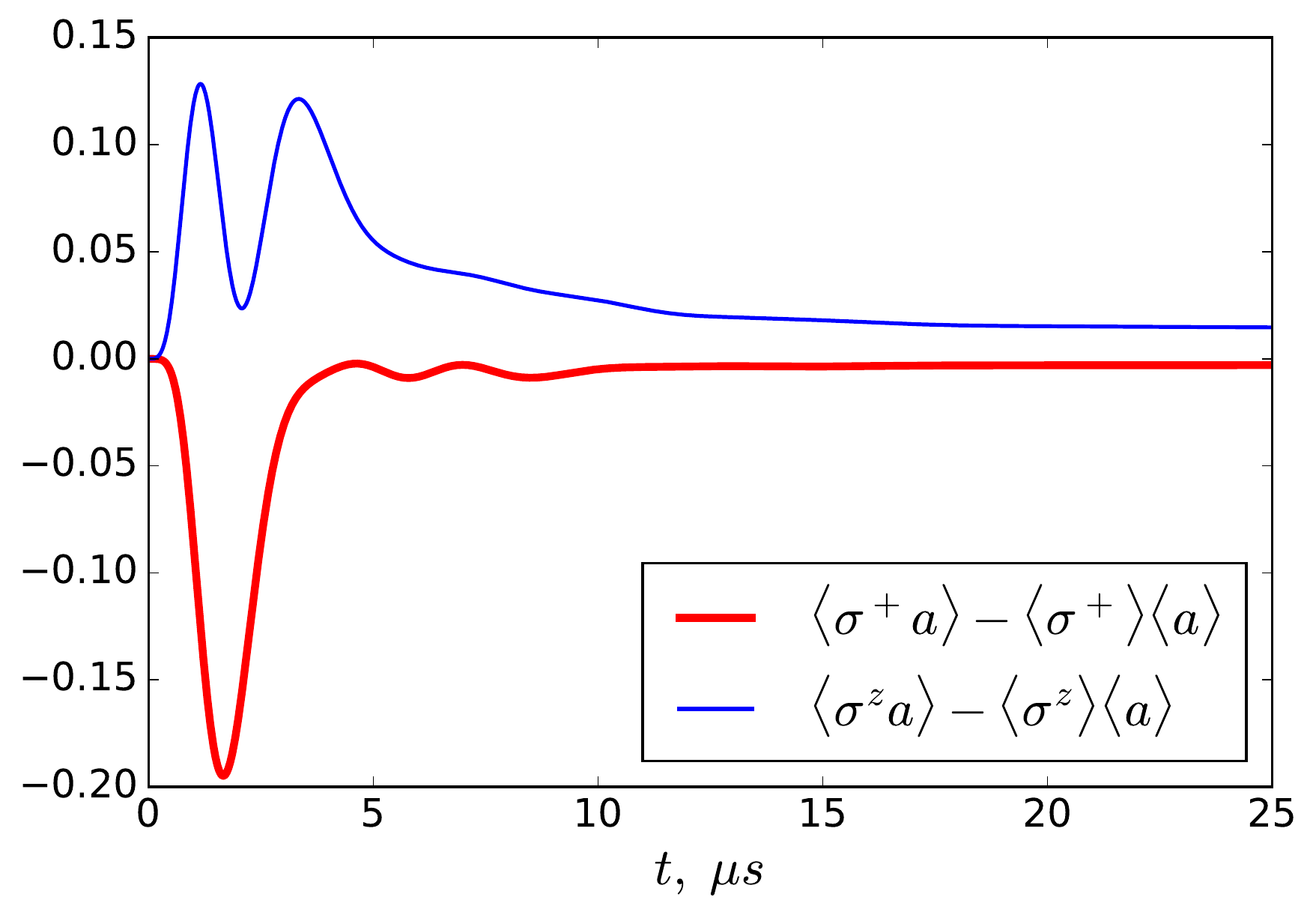}
	\caption{Time evolution of correlations $ \langle \delta a\delta\sigma^{+}\rangle$ and $ \langle \delta a\delta\sigma^{z}\rangle$ extracted from the solution of the Lindblad equation at $f=g$.} \label{result:nonstationary:spl-a}
\end{figure}

The second result of our paper is that quantum corrections $\langle \delta a\delta\sigma^{z,+}\rangle$ play a significant role in the non-stationary dynamics of the quantum interface even at drivings less than the steady state threshold $f^*$. This is demonstrated via  time evolution of $n_{ph}(t)$ and $n_q(t)$ after the moment $t=0$ when the external driving is suddenly switched on. The threshold value, observed for the steady state regime, is estimated as $f\approx 1.5 g$ for our parameters of the system. We set the after-quench value of the driving at the smaller value $f=g$.  Figures \ref{result:nonstationary:ns}, \ref{result:nonstationary:nph} and \ref{result:nonstationary:spl-a} demonstrate the distinctions between the qubit and photon occupation number dynamics obtained from non-stationary solutions of the Maxwell-Bloch (\ref{a},\ref{sigma_minus},\ref{sigma-z}) and Lindblad equation (\ref{lindblad}).
\begin{figure}[h]	\includegraphics[width=\linewidth]{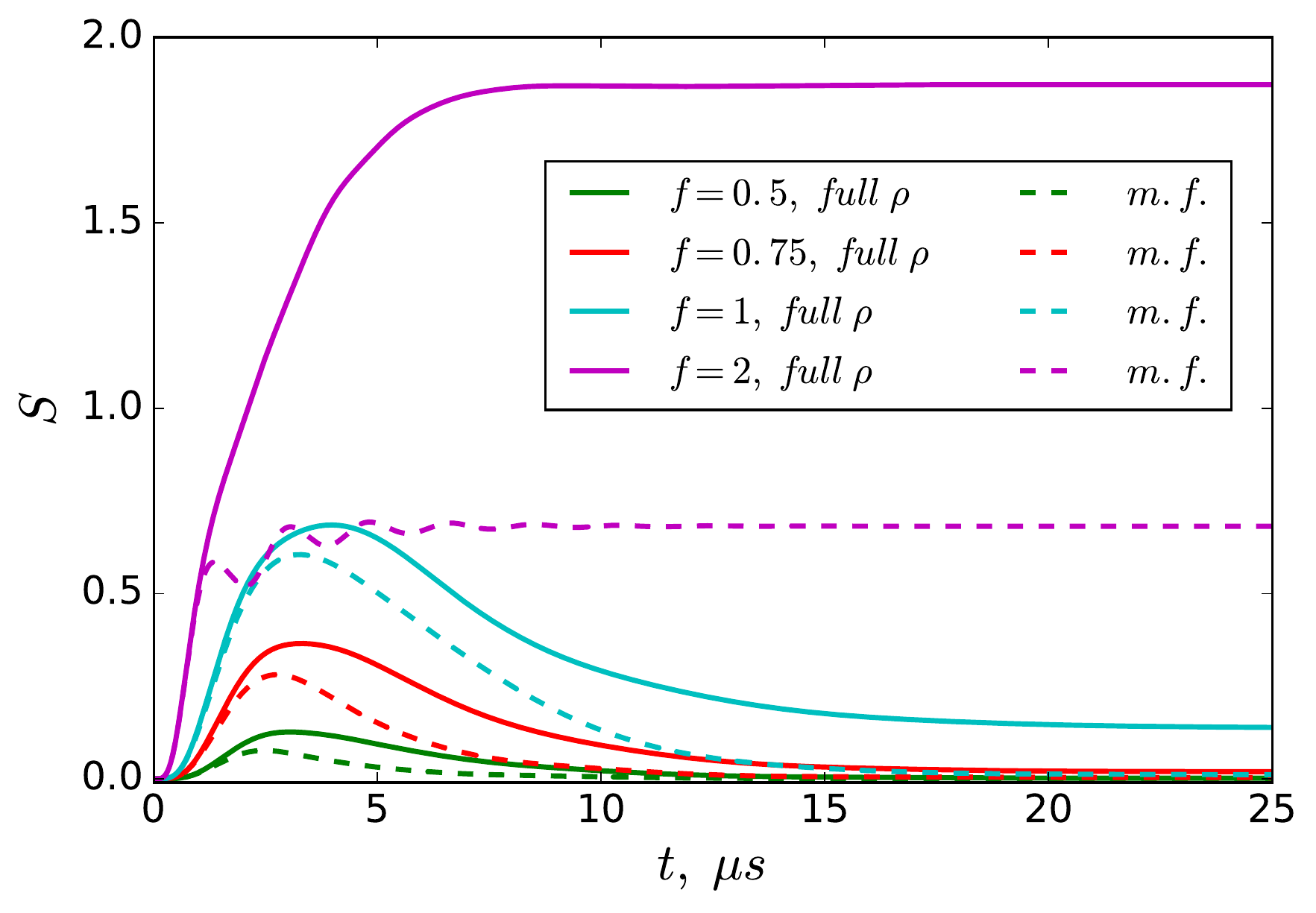}
	\caption{Time evolution of entropy  $S$  at different amplitudes of the driving $f$.} \label{result:nonstationary:s}
\end{figure}

In Figure \ref{result:nonstationary:s} we present the results for von Neumann entropy as function of time at different values of the driving $f$. 
%It is seen from both of Figures \ref{result:stationary:s} and \ref{result:nonstationary:s} that the entropy saturates at $f>f^*\approx 1.5 g$, i.e. exceeding the threshold value. 
We observe a strong difference in values entropy found from solving of Lindblad (solid curves) and mean-field (dashed curves) equations. For $f>f^*$, the entropy grows almost monotonically, until the saturation at the steady-state value. Contrary, for $f<f^*$ there is a pronounced maximum at $t\approx 4 \mu s$. The peak is present and the full-$\rho$ result is different from the mean-field one even for a weak driving $f=0.1 g$, although the steady-state entropy is almost vanished for much larger $f=0.75 g$. This indicates an emergent entanglement between qubit and photon mode of the quantum interface being switched.

\section{Conclusions}
We have studied  the response of a dissipative  hybrid qubit-cavity system to the applied strong driving field, having in mind the future possible realization of quantum operations in superconducting quantum metamaterials. We demonstrated  that for the case studied the many-body effects (or, equally, the  entanglement between the qubit and photon excitations) are important and that the system  cannot be treated by means of a mean-field approximation.
This is shown from a comparison of  analytical steady state solution of the standard Maxwell-Bloch equations and  numerical simulations based on Lindblad equation on the many-body density matrix. Speaking more concretely, we have shown that mean-field approach, where the density matrix of the system can be represented via direct product of isolated qubit and photon ones $\rho_{mf}=\rho_{ph} \otimes \rho_q$, provides a good steady state solution up to certain threshold $f^*$ but at $f>f^*$ the strong discrepancy from the many-body result is observed.  It is related with a growing value of quantum  correlations between fluctuations of qubit and photons fields which start to play a significant role in behavior of the system. 
We show in our analysis that at large enough coupling energy between cavity and qubit modes the solution of Maxwell-Bloch equations reveals an artifact  being a hysteresis in number of photons as function of the driving amplitude in vicinity of the threshold $f^*$. Such a hysteresis has not been observed in the full density matrix solution. Also we have studied an effect of the non-adiabatic  switching of the driving and show that there is a difference between mean-field and the density matrix solutions even for the drivings weaker than  the  steady state threshold $f^*$.

Our findings demonstrate quantitative limitations of standard mean-field description and show the crossover between the classical and many-body quantum regimes. In the classical regime, the qubit virtually acts as a linear 
(Gaussian) degree of freedom; this regime cannot reveal a difference between the quantum and linear-optical metamaterials. When the two-level nature of the qubit plays an essential role, its entanglement with the cavity mode is also large and should be accounted. We also point out that the effect of correlations is revealed while the number of photons in the cavity mode is not small and one could naively expect that the cavity operates in a classical regime.
  In our solutions we have used  parameters relevant for contemporary metamaterials involving highly anharmonic flux qubits, and we expect that the obtained results will find an application in realization of quantum gates in superconducting quantum circuits and metamaterials.

 \section{Acknowledgments}
 Authors thank Yuriy E. Lozovik, Andrey A. Elistratov, Evgeny S. Andrianov and Kirill V. Shulga for fruitful discussions.  The study was funded by the Russian Science Foundation (grant No. 16-12-00095).


\begin{thebibliography}{<num>}
	\bibitem{Astafiev}  O. Astafiev, A. M. Zagoskin, A. A. Abdumalikov Jr.,   Yu. A. Pashkin, T. Yamamoto, K.  Inomata, Y.  Nakamura, and J. S.  Tsai,    Science  {\bf 327}, 840  (2010).
	\bibitem{Macha}  P. Macha, G. Oelsner, J.-M. Reiner,  M. Marthaler, S. Andr\'e, G. Sch\"on, U. H\"ubner,  H.-G. Meyer,  E. Il'ichev, and   A.V. Ustinov,  Nature  Commun. {\bf 5}, 5146 (2014).
	\bibitem{Rakhmanov}  A. L. Rakhmanov, A. M.  Zagoskin,   S. Savel'ev, and F.  Nori,    Phys. Rev. B  {\bf 77}, 144507 (2008).
	\bibitem{Fistul}P. A. Volkov and   M. V. Fistul,  Phys. Rev. B  {\bf 89}, 054507 (2014).
	\bibitem{ZKF}  N. I. Zheludev and Y. S. Kivshar, Nat.
	Mater. {\bf 11}, 917-924 (2012).
	\bibitem{SMRU}  D. S. Shapiro, P. Macha, A. N. Rubtsov,  and A. V.  Ustinov,   Photonics {\bf 2}  (2), 449-458 (2015).
	\bibitem{Brandes} T. Brandes, Physics Reports {\bf 408}, 315 (2005)
	\bibitem{Zou} L. J. Zou, D. Marcos, S. Diehl, S. Putz, J. Schmiedmayer, J. Majer, and P. Rabl, Phys. Rev. Lett. {\bf 113}, 023603 (2014)
	
	\bibitem{nv-centers} S. Putz, D. O. Krimer, R. Ams\" uss, A. Valookaran, T. N\" obauer, J. Schmiedmayer, S. Rotter and J. Majer, Nature Physics {\bf 10}, 720-724 (2014).
	\bibitem{nv-centers-1}  K. Sandner, H. Ritsch, R. Ams\" uss, Ch. Koller, T. N\" obauer, S. Putz, J. Schmiedmayer, and J. Majer,  Phys. Rev. A {\bf 85}, 053806 (2012).
	\bibitem{nv-centers-0} M. V. G. Dutt, L. Childress, L. Jiang, E. Togan, J. Maze, F. Jelezko, A. S. Zibrov, P. R. Hemmer, and M. D. Lukin, Science
	{\bf 316}, 1312 (2007).
	
	\bibitem{Morton}     J. J. L. Morton, A. M. Tyryshkin, R. M. Brown, S. Shankar, B. W. Lovett, A. Ardavan, T. Schenkel, E. E. Haller, J. W. Ager, and  S. A. Lyon, Nature {\bf 455}, 1085-1088 (2008).
	\bibitem{Schuster} D. I. Schuster,  %{\em et al.},
	A. P. Sears, E. Ginossar, L. DiCarlo, L. Frunzio, J. J. L. Morton, H. Wu, G. A. D. Briggs,
	B. B. Buckley, D. D. Awschalom, and R. J. Schoelkopf,
	Phys. Rev. Lett. {\bf 105}, 140501 (2010).
	
	\bibitem{MSS} Y. Makhlin, G. Sch\" on, and A. Shnirman, Rev. Mod. Phys. {\bf 73}, 357-400 (2001).
	\bibitem{Orlando}  T. P. Orlando,   J. E.  Mooij, L. Tian,    C. H. van der Wal, L. S. Levitov, S. Lloyd, and  J. J. Mazo,  Phys. Rev. B {\bf 60}, 15398 (1999).
	\bibitem{mooij} J. E. Mooij,   T. P. Orlando,  L. Levitov,    L. Tian,   C. H. van der Wal, and S. Lloyd,   Science {\bf 285}, 1036 (1999).
	
	\bibitem{Clarke} J. Clarke and F. K. Wilhelm, Nature {\bf 453}, 1031-1042 (2008).
	\bibitem{DiCarlo}   L. DiCarlo,  J. M. Chow,  J. M. Gambetta, L. S.  Bishop, B. R.  Johnson, D. I.  Schuster,  J.  Majer, A.  Blais, L.  Frunzio, S. M.  Girvin, R. J.  Schoelkopf,    Nature {\bf 460}, 240 (2009).
	\bibitem{Nation}  P. D. Nation,  J. R. Johansson, M. P. Blencowe, and F.  Nori,     Rev. Mod. Phys. {\bf 84}, 1  (2012).
	
	
	
	
	
	\bibitem{Carmichael} H. J. Carmichael, {\it Statistical Methods in Quantum Optics 1}, (Springer-Verlag Berlin Heidelberg, 1999).
	
	
	\bibitem{SavageCarmichael} C. M. Savage and H. J. Carmichael, IEEE Journal of Quantum Electronics, vol. {\bf 24}, No. 8 (1988).
	

	
%\bibitem{Blais} A. Blais, R.-S. Huang, A. Wallraff,  S. M. Girvin, and R. J. Schoelkopf,     Phys. Rev. A  {\bf 69}, 062320 (2004).
%\bibitem{YouNori}J. Q. You and  F. Nori,   Nature {\bf 474},  589-597 (2011).
%\bibitem{Stanwix} P. L. Stanwix,  L. M. Pham, J. R. Maze, D. Le Sage, T. K. Yeung, P. Cappellaro, P. R. Hemmer, A. Yacoby, M. D. Lukin, and R. L. Walsworth, Phys. Rev. B {\bf 82}, 201201 (2010).
%	\bibitem{Wesenberg0}J. H. Wesenberg, A. Ardavan, G. A. D. Briggs, J. J. L. Morton,
%	R. J. Schoelkopf, D. I. Schuster, and K. M\o lmer, Phys. Rev. Lett.
%{\bf 103}, 070502 (2009).
%	\bibitem{Wesenberg}J. H. Wesenberg, Z. Kurucz, and K. M\o lmer, Phys. Rev. A {\bf 83}, 023826 (2011)
%	\bibitem{Grezes} C. Grezes, B. Julsgaard, Y. Kubo, M. Stern, T. Umeda, J. Isoya, H. Sumiya, H. Abe, S. Onoda, T. Ohshima, V. Jacques, J. Esteve, D. Vion, D. Esteve, K. M\o lmer, and P. Bertet, Phys. Rev. X {\bf 4}, 021049 (2014).
%\bibitem {Wu} H. Wu, R. E. George, J. H. Wesenberg, K. M\o lmer, D. I. Schuster, R. J. Schoelkopf, K. M. Itoh, A. Ardavan, J. J. L. Morton, and G. A. D. Briggs,  Phys. Rev. Lett. {\bf 105}, 140503 (2010).
%	\bibitem{Moelmer}  B. Julsgaard and K. M\o lmer Phys. Rev. A {\bf 88}, 062324 (2013).
%	\bibitem{spinref-1} N. W. Carlson, L. J. Rothberg, A. G. Yodh, W. R. Babbitt, and
%	T. W. Mossberg, Opt. Lett. {\bf 8}, 483 (1983).
%	\bibitem{spinref-2} H. Lin, T. Wang, and T. W. Mossberg, Opt. Lett. {\bf 20}, 1658
%	(1995).
%	\bibitem{ASRG} M. Afzelius, C. Simon, H. de Riedmatten, and N. Gisin,
%	Phys. Rev. A {\bf 79}, 052329 (2009).
%	\bibitem{rangelov} A. A. Rangelov, N. V. Vitanov, L. P. Yatsenko, B. W. Shore, T. Halfmann, and K. Bergmann, Phys. Rev. A {\bf 72}, 053403 (2005).
%	\bibitem{schoenfeldt}  J.-H. Sch\" onfeldt, J. Twamley, and S. Rebi\' c, Phys. Rev. A {\bf 80}, 043401 (2009).
%	\bibitem{bergmann} K. Bergmann, H. Theuer, and B. W. Shore, Rev. Mod. Phys. {\bf 70}, 1003 (1998).
%	\bibitem{remizov} S. V. Remizov, D. S. Shapiro, and A. N. Rubtsov,  Synchronization of qubit ensemble under optimized $\pi$-pulse driving, Phys. Rev. A {\bf 92},  053814 (2015).

\end{thebibliography}
\end{document}